# $^{23}$Na and $^1$H NMR Relaxometry of Shale at High Magnetic Field

*Donghan Yang and Ravinath Kausik\**

Schlumberger-Doll Research Center, One Hampshire Street, Cambridge, MA 02139 USA

**ABSTRACT**:

Formation evaluation of unconventional reservoirs is challenging due to the coexistence of different phases such as kerogen, bitumen, movable and bound light hydrocarbon and water. Current low-frequency (0.05 T) nuclear magnetic resonance (NMR) laboratory and logging methods are incapable of quantitatively separating the different phases. We demonstrate the utility of high-field (9-T) NMR 2D $T_1$-$T_2$ measurements for separating hydrocarbon and the clay-interacting aqueous phases in shale based on the difference in the frequency dependence of the spin-lattice relaxation time. Furthermore, we demonstrate $^{23}$Na NMR as a promising complementary technique to conventional $^1$H NMR for shale fluid typing, taking advantage of the fact that sodium ions are only present in the aqueous phase. We validate high-field (9 T) $^{23}$Na-$^1$H NMR relaxometry for assessing brine-filled porosity and brine salinity in various porous materials, including porous glass, conventional rocks, clays, and shale, and apply it for differentiating hydrocarbon versus aqueous components and also the clay-associated versus free water in Eagle Ford shale cores. This work lays the groundwork for developing future downhole $^{23}$Na-$^1$H NMR logging techniques.

# 1. INTRODUCTION

The formation quality of a tight oil shale play is reflected by the Reservoir Producibility Index (RPI), a function of movable light oil, bound light oil, bitumen, kerogen, free water, and bound water.[1] Therefore, it is important to measure the component-specific porosity of all coexisting components in the reservoir for the identification of "sweet spots" for landing horizontal wells. Nuclear magnetic resonance (NMR) has been extensively used for analyzing fluid properties, such as porosity, permeability, wettability, and diffusivity, in porous materials ranging from biological microstructures to oil/gas reservoirs.[2-12] Historically, commercial NMR well-logging tools primarily utilize low Larmor frequency (< 2 MHz) proton ($^1H$) NMR for fluid characterization in conventional and unconventional reservoirs, and the typical modalities include NMR signal intensity, relaxometry, and diffusometry, such as 1D $T_2$, 2D $T_1$-$T_2$, and $D$-$T_1$-$T_2$ distributions ($T_1$ and $T_2$ being the spin-lattice and spin-spin relaxation times, respectively, and $D$ the apparent diffusion coefficient).[7, 13-19] As a quantitative method, the $^1H$ NMR signal intensity is proportional to the number of $^1H$ nuclei in reservoir fluids, and thus NMR signal can be converted into the quantity (mass and volume) of hydrocarbons, using the relevant physical and chemical properties of the fluids (*e.g.*, type of hydrocarbon and hydrogen index). Furthermore, NMR relaxometry and diffusometry data contain the information about not only the fluids (intrinsic $T_1$ and $T_2$ and apparent diffusion coefficient)[19] but also the confining systems, namely, surface relaxivity (e.g., surface composition and pore wettability) and pore geometry (e.g., pore size/shape, tortuosity, and permeability). Therefore, both fluid typing and pore-size probing can be achieved through NMR relaxometry/diffusometry experiments in the reservoir.

In unconventional tight oil shales, the $^1$H NMR $T_1/T_2$ distributions for bitumen and bound water tend to overlap, thus leaving insufficient contrast for separating these different reservoir fluids, especially at low Larmor frequencies.[20] This impedes the task of differentiating the porosity of hydrocarbon and aqueous phases. In this paper, we show how high-field (9 T; 400 MHz for $^1$H) NMR relaxometry can take advantage of the difference in the frequency dependence of spin-lattice relaxation time and thus better separate the different components in tight oil shales.

We further discuss the application of a parallel NMR marker, $^{23}$Na, for enhancing the fluid typing functionality of NMR. Sodium is generally the most abundant cation in reservoir fluids in unconventional plays as demonstrated by a survey of the production water from 391 wells in the Bakken formation (http://eerscmap.usgs.gov/pwapp), which reveals the average Na$^+$ concentration to be 3.16 mol/L (salinity ~200 ppk; ppk defined as g NaCl / kg H$_2$O), suggesting a salinity as high as that in the Dead Sea. This abundance of sodium nuclei in the reservoir water makes NMR measurements feasible. Furthermore, sodium is only present in the aqueous compartment (i.e., brine) in tight oil shales, unlike proton ($^1$H) which is present in both hydrocarbon and water. Therefore, $^{23}$Na NMR serves as a probe for the aqueous components (e.g., bound and free water) in shales, helping separate the signals and thus the porosities of hydrocarbons and brine. In addition to fluid typing and porosity measurement, the quantitative analysis of downhole Na$^+$ concentration might also benefit conventional $^1$H NMR logging by providing important correction information for hydrogen index and logging tool performance.[21]

In this study, we discuss the application of high-field NMR $^1$H relaxometry and the validation of $^{23}$Na NMR as a parallel modality for quantitative fluid typing in tight oil shale. We demonstrate the application of $^1$H 2D $T_1$-$T_2$ maps at high Larmor frequency (400 MHz) for identifying different components in both native and fluid-resaturated Eagle Ford shale. Additionally, we demonstrate a

abstract$^{23}$Na NMR protocol for brine quantification and evaluate it in controlled synthetic porous glass (CPG), conventional rocks, and clays. The combined $^{23}$Na and $^{1}$H NMR methodology is shown to enable quantification of the brine-filled porosity in shale and for separating bound and free fluids based on the mobility of sodium nuclei in brine. We further show that the NaCl salinity of brine confined within porous materials can be readily derived using $^{23}$Na-$^{1}$H-combined NMR. We also discuss how further studies of $^{23}$Na surface relaxivity can help improve the characterization of unconventional shale samples.

## 2. THEORETICAL BASIS

$T_1$ and $T_2$ relaxation times are sensitive to motions at the Larmor frequency and low or zero frequency, respectively. Therefore, $T_1$-$T_2$ measurements are sensitive to the inherent molecular mobility of the different species. Based on this principle, the $T_1$-$T_2$ map can be used to separate different fluids. This methodology has been applied at 2-MHz Larmor frequency to separate different components in shale samples.[20, 22] In particular, the bitumen and bound water components can be separated from the oil in organic porosity and the fluids in inorganic porosity based on the $T_2$ values and $T_1/T_2$ ratios in the 2D $T_1$-$T_2$ maps.[22] One of the biggest shortcomings of these low Larmor frequency measurements is their inability to separate bitumen (viscous heavy oil) from bound water, which is because at 2 MHz, both these components generally have comparable $T_2$ values, and adequate contrast in $T_1/T_2$ ratios might not always be available.[22]

High-field (high Larmor frequency) NMR is a potential candidate to separate these two components for two reasons. First, these measurements have a higher sensitivity to motions ranging from the higher Larmor frequency to zero frequency. Second, they can take advantage of the fact that $T_1$ of heavy oil (bitumen) has $1/\omega$ frequency dependence while that of bound water has $1/\log(\omega)$ frequency dependence.[23] Therefore, at high-enough Larmor frequencies, the $T_1$ values of these two components will be different, thus enabling their differentiation. Additionally, the kerogen component is also visible in high frequency measurements due to the possibility of application of shorter echo times.

The $^{23}$Na nucleus is the only stable isotope of sodium. It has a gyromagnetic ratio of 11.26 MHz/T, which is 26.4% of that of $^1$H.[24] In bulk aqueous solution of NaCl, the $T_1$ of $^{23}$Na is only ~3% of the $T_1$ of $^1$H,[25, 26] which assures a significantly faster signal acquisition cycle (typically with the duration of 3–5 $T_1$) compared with $^1$H experiments. Accordingly, within a given experimental time window, $^{23}$Na NMR allows as many as 33 times more signal averages than $^1$H NMR, greatly ameliorating the difficulty of the relatively low NMR signal sensitivity for $^{23}$Na (9.25% of the sensitivity of $^1$H).[24] A summary of the relevant NMR properties of $^{23}$Na and $^1$H is given in Table 1.

$^{23}$Na has a spin of 3/2, so its NMR relaxation is primarily driven by the interaction between the nuclear electric quadruple moment and the local electric field gradient (EFG), i.e., through electric quadrupole coupling.[27, 28] $^{23}$Na relaxation times[28] for longitudinal (spin-lattice) relaxation are

$$M_z(t) = M_z(\infty) - [M_z(\infty) - M_z(0)]\left[\frac{1}{5}\exp(-R_{1a}t) + \frac{4}{5}\exp(-R_{1b}t)\right] \quad (1a)$$

$$R_{1a} = Aj_1(\omega_0) \quad (1b)$$

$$R_{1b} = Aj_2(2\omega_0) \quad (1c)$$

and those for transverse (spin-spin) relaxation are

$$M_{xy}(t) = M_{xy}(0)\left[\frac{3}{5}\exp(-R_{2a}t) + \frac{2}{5}\exp(-R_{2b}t)\right] \quad (2a)$$

$$R_{2a} = \frac{1}{2}A[j_0(0) + j_1(\omega_0)] \quad (2b)$$

$$R_{2b} = \frac{1}{2}A[j_1(\omega_0) + j_2(2\omega_0)] \quad (2c)$$

In Eqs. (1) and (2), $M(t)$, $M(0)$, and $M(\infty)$ are the time-dependent, initial, and equilibrium magnetizations, respectively, and subscripts $z$ and $xy$ denote longitudinal and transverse magnetizations, respectively. $R_1$ is the longitudinal relaxation rate constant ($T_1 = 1/R_1$) and $R_2$ is the transverse relaxation rate constant ($T_2 = 1/R_2$). Note the rate constants are determined by coefficient $A$ and the reduced spectral densities $j_n(n\omega_0)$ ($n = 0$, 1, and 2). The coefficient $A$ is a function of both the nuclear electric quadrupole moment and the local EFG and depicts the intensity of electric quadrupole coupling.[28] The reduced spectral densities have a Lorentzian form for the bulk fluids and are given as:

$$j_n(n\omega_0) = \frac{\tau_c}{1 + (n\omega_0\tau_c)^2} \quad (3)$$

where $\omega_0$ is the Larmor frequency of $^{23}$Na, and $\tau_c$ is the quadruple coupling correlation time, which describes the time scale of the fluctuation of the electric quadrupole coupling.[28] In the general analysis, as suggested by Eq. 1, both the longitudinal and transverse relaxations of $^{23}$Na follows a bi-exponential form. This is significantly different from the mono-exponential relaxations for $^1$H, which is spin-1/2 nucleus and does not experience electric quadrupole coupling.

In a dilute aqueous solution of NaCl as considered in this paper, the electric dipole moments of water molecules are the chief contributors to the local EFG experienced by sodium ions[28] and the rapid motion of water molecules and ions creates an isotropic ionic environment for $^{23}$Na nuclei.[27] For bulk solutions, the short correlation time ($\tau_c \sim$ 3–10 ps)[28] leads to an averaging of EFG, and the "extreme narrowing condition",[24] $\omega_0\tau_c \ll 1$, is fulfilled.[25, 29] Therefore, the intrinsic bi-exponential relaxations of $^{23}$Na become mono-exponential with $R_{ia} = R_{ib}$ ($i$ = 1 or 2) as suggested by Eqs. 1 through 3. On the other hand, in anisotropic systems, e.g., model biopolymer gels[28] and dense Laponite clay suspensions[30, 31], bi-exponential relaxation behavior is exhibited due to the relatively slow modulation of electric quadrupole coupling (relatively long $\tau_c$).

The NMR relaxation of $^{23}$Na in model porous materials[29, 32] and conventional rocks[24] has been shown to be mainly mono-exponential as that observed in bulk NaCl solutions[25]. In addition, $^{23}$Na relaxation times in model porous materials reflect pore-size information[29] just as in the case of $^1$H NMR[19]:

$$\frac{1}{T_{i,\text{pore}}} = \frac{1}{T_{i,\text{bulk}}} + \frac{S}{V}\rho_i \qquad (i = 1 \text{ or } 2) \tag{4}$$

where $T_{i,\text{pore}}$ is the relaxation time in pores, $T_{i,\text{bulk}}$ is the concentration-dependent relaxation time in bulk NaCl solutions, $S/V$ is the surface-area-to-volume ratio of the pores, and $\rho_i$ is the surface relaxivity. In low field systems where internal gradients are negligible, when $T_{i,\text{bulk}}$ and $\rho_i$ are known, pore-size information, i.e., $S/V$, can be retrieved by measuring $T_{i,\text{pore}}$. A more complicated scenario can occur in extremely small pores formed by clay and saturated with brine, where a larger fraction of sodium ions are immobile close to the surface layer (as Na$^+$ is the neutralizing counterion at the surface of many kinds of clay materials)[30, 31] and are thus subject to higher anisotropic EFG.[28, 29] In this case, motional averaging of electric quadrupole coupling is ineffective so bi-exponential $^{23}$Na relaxations emerge, thereby making the interpretation of $^{23}$Na relaxation distributions more complicated.

**Table 1. NMR properties of $^{23}$Na and $^{1}$H.**

| Nucleus | Spin | Natural abundance (%) | Gyromagnetic ratio (MHz/T) | Larmor frequency at 9.4 T (MHz) | Relative sensitivity |
| --- | --- | --- | --- | --- | --- |
| $^{1}$H | 1/2 | 99.98 | 42.58 | 400.2 | 1.0 |
| $^{23}$Na | 3/2 | 100.0 | 11.26 | 105.8 | 0.0925 |

## 3. EXPERIMENTAL SECTION

### 3.1. Sample Preparation.

*Reference solutions for NMR signal calibration*

Reference solutions for $^{23}$Na NMR were brines prepared at various salinities using dry NaCl salt and deionized water. Reference solutions for $^1$H NMR were prepared by diluting deionized water (H$_2$O) using deuterium oxide (D$_2$O) at various H$_2$O/D$_2$O ratios. At 9-T magnetic field, the volume of reference solution was kept at 100 μL for each signal-calibration point whereas the quantity of NaCl ranged from 0.1 to 1.8 mg (for $^{23}$Na calibration) and that of H$_2$O ranged from 2 to 14 mg (for $^1$H calibration).

*Porous glass and conventional rocks*

For high-field (9-T) NMR experiments, one type of model porous Vycor glass (Corning, NY, USA) and three types of conventional rock, namely, Austin Chalk, Edward Limestone, and Berea Sandstone 500 (Berea 500), were studied. All samples were cut into cylindrical plugs with length of ~10–15 mm, diameter of ~3 mm, and total volume of ~70–100 mm$^3$. To prepare brine-saturated samples, sample plugs were first kept at 100° C in an oven for 1–2 days for initial dehydration and then moved into a laboratory-built vacuum/pressure chamber, where the samples were maintained under vacuum at −1 bar and room temperature for 2 days for further dehydration. After the second round dehydration, brine (32 ppk) was injected into the chamber at +70-bar inner pressure, and the samples were maintained under this condition for another 2 days. The mass increase in a sample plug from its dry state (after initial dehydration) to its wet state (after pressurized saturation) was used, along with plug volume and brine density, to calculate the brine-filled porosity of the sample (also referred to as the mass-difference porosity, or MD porosity, in Tables 2-4). The saturated glass and rock samples were then used for high-field $^{23}$Na and $^1$H NMR experiments. $^1$H NMR at low magnetic field (0.05 T) was also performed as a reference for the high-field experiments. For low-field $^1$H NMR measurements, the same types of porous glass and conventional rocks were prepared in larger dimensions (5 cm in diameter and length) and saturated with 32-ppk brine using the same procedure.

*Clays*

For high-field NMR experiments, pure quartz and three types of clay, namely, illite, bentonite, and smectite, were separately mixed with 90-ppk brine using the following procedure. First, brine (90 ppk; ~20–30 µL) was combined with dry quartz or clay powder in a 5-mm NMR tube, resulting in a wet mixture. To prevent water from evaporating, additional dry powder of the same type of material (quartz or clay) was added and the sample was sealed with a piece of Teflon tape. The sample was then tightly compacted at the bottom of the NMR tube using a plunger. The total length of the sample did not exceed 15 mm, which was within the sensitivity volume of the NMR probe.

*Shale*

For high-field NMR experiments, native-state Eagle Ford shale samples from three different depths were cut into small yet well-defined pieces (~10–20 mm$^3$ per piece). A portion of each native-state sample was packed into a 5-mm NMR tube for control study. The remaining samples were saturated with 110-ppk brine using the same procedure for preparing porous glass and conventional rocks except that the dehydration steps were skipped. The mass increase in each sample upon brine saturation was used, along with the mass and density of the native-state samples, to calculate the mass-difference porosity.

### 3.2. NMR Experiments

High-field NMR experiments were performed on a 9-T Bruker system with a console working at 400 MHz for $^1$H and 106 MHz for $^{23}$Na (Bruker, Billerica, MA, USA). Three NMR pulse sequences were used for both $^1$H and $^{23}$Na studies: free induction decay (FID; Fig. 1A), Car-Purcell-Meiboom-Gill (CPMG; Fig. 1B), and inversion-recovery CPMG (IR-CPMG; Fig. 1C). FID data were acquired with a sampling dwell time of 1 μs. The echo time in CPMG and IR-CPMG acquisitions was typically 49 μs for $^1$H and 54 μs for $^{23}$Na. The recycle delay in all experiments was kept at five times the longest $T_1$ in each sample. In IR-CPMG experiments, 32 logarithmically spaced inversion times (*TI*) were used for measuring $T_1$ distribution. NMR tubes made of synthetic quartz (Wilmad-LabGlass, Vineland, NJ, USA) were used for all experiments that involved $^{23}$Na measurement to minimize $^{23}$Na background signals, which exist in normal glass tubes.

Low-field $^1$H NMR experiments were used as reference for high-field studies in Vycor glass and rocks. Low-field experiments were performed on a 0.05-T Geospec system (2 MHz for $^1$H; Oxford Instruments, Oxford, UK). CPMG and IR-CPMG data were collected using an echo time of 100 μs.

NMR 1D $T_2$ and 2D $T_1$-$T_2$ distributions were derived from CPMG and IR-CPMG data, respectively, using a laboratory-developed inverse-Laplace-transform MATLAB program (MathWorks, Natick, MA, USA).[33] NMR 1D $T_1$ distributions were extracted from the projection of $T_1$-$T_2$ maps onto the $T_1$ dimension.

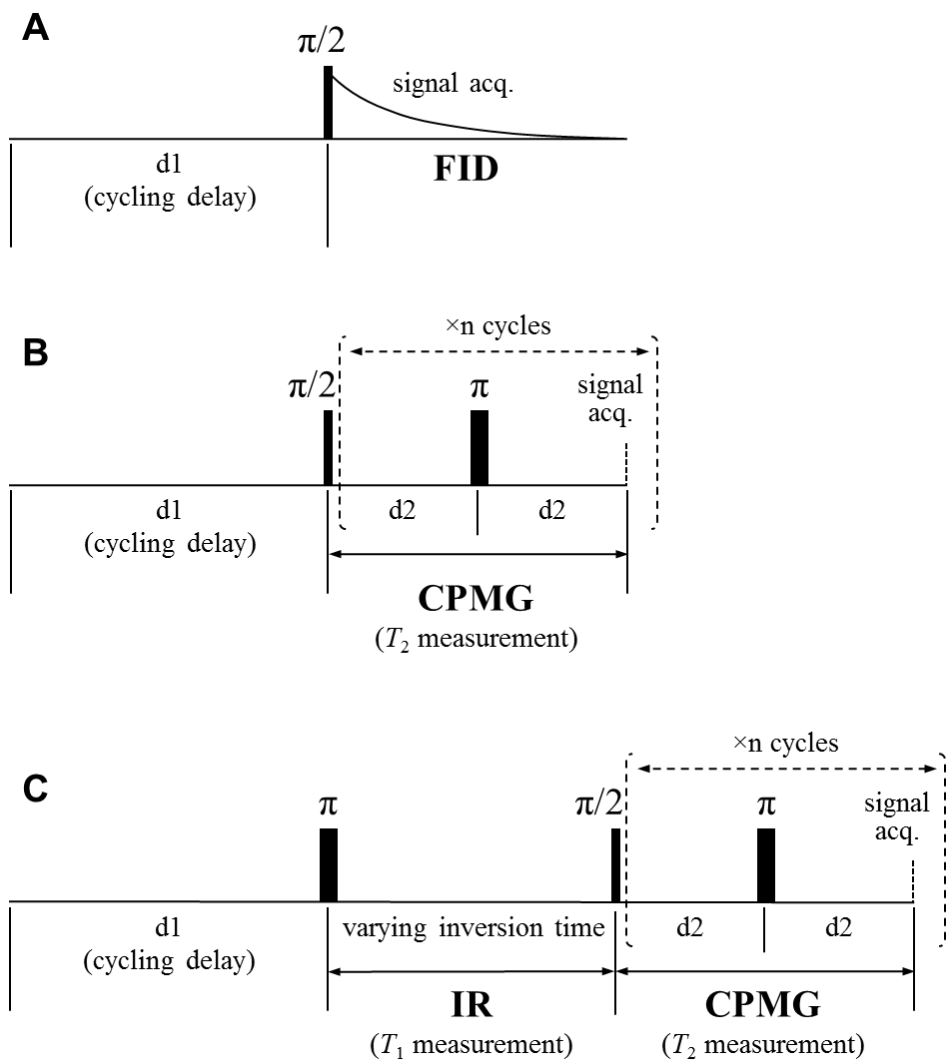

**Figure 1.** NMR pulse sequences. A) Free induction decay (FID); B) Carr-Purcell-Meiboom-Gill (CPMG); and C) inversion-recovery (IR) CPMG.

## 4. RESULTS AND DISCUSSION

We present the results in three subsections below, according to the three groups of samples that were studied. Within each subsection, the experimental results are discussed with an emphasis on $^1$H- and $^{23}$Na-NMR based 1) fluid typing and 2) relaxometry.

To quantify the experimental results, NMR signal calibration was established for both $^1$H and $^{23}$Na channels by determining the linear correlation between the observed NMR signal and the quantity of the target molecule in the reference solutions. The target molecule for $^1$H NMR was $H_2O$ in $H_2O/D_2O$ solution whereas that for $^{23}$Na NMR was NaCl in NaCl/$H_2O$ solution (brine). The least squares fitting of the calibration lines (CPMG acquisition) is shown in Fig. 2 for $^1$H ($r^2 = 0.999$) and $^{23}$Na ($r^2 = 0.995$), respectively, with $r$ being the correlation coefficient. The calibration line for $^1$H measurements does not pass through the origin, due to a small background $^1$H signal. This background signal is accounted for during quantitative measurements.

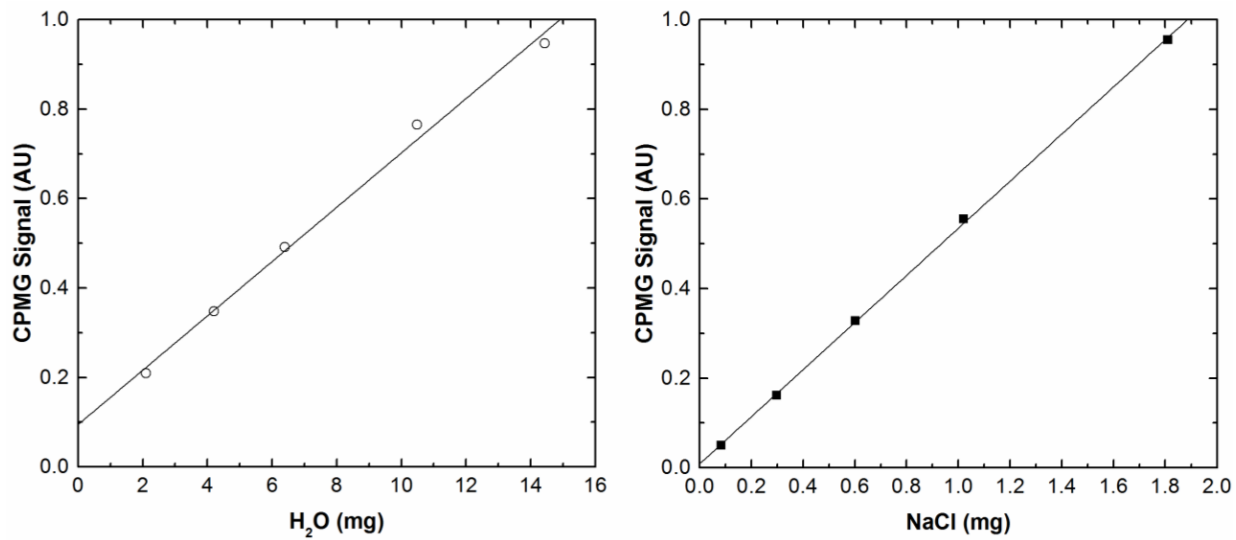

**Figure 2.** NMR signal calibration reference. CPMG signals are plotted against the mass of $H_2O$ (open circles, left plot) and NaCl (filled squares, right plot), respectively, in bulk reference NaCl solutions. Least squares linear fitting results are shown by solid lines in both plots.

## 4.1. Porous Glass and Conventional Rocks

The NMR $T_2$ distributions for brine-saturated Vycor glass, Austin Chalk, and Edward Limestone were converted into brine-filled porosities via $^1$H and $^{23}$Na signal calibration because the salinity of the saturating brine (32 ppk) was known. The porosities separately determined by $^1$H and $^{23}$Na measurements (referred to as $^1$H- and $^{23}$Na-porosity, respectively) are shown in Table 2. Both types of porosity match well with the mass-difference (MD) porosity, which was based on the mass increase in the samples upon brine saturation. The uncertainty in $^{23}$Na-porosity is relatively higher than that in $^1$H-porosity because the amount of NaCl in these samples (~0.5 mg) was much smaller than that of $H_2O$ (~16 mg). In Edward Limestone, $^1$H-porosity (18.7 ± 0.3 p.u.) is slightly lower than the corresponding MD porosity (22.3 ± 0.2 p.u.), suggesting a difference of ~2.4 mg in the mass of water arising from water evaporation while packing the sample into the NMR tube. In contrast, the $^{23}$Na-porosity (22 ± 2 p.u.) remains identical to the MD porosity because water evaporation only leaves behind a more concentrated brine. The last column in Table 2 shows the NMR-derived brine salinity, which was calculated using the mass of NaCl and $H_2O$ obtained from NMR measurements. The NMR-derived salinities agree well with the actual brine salinity (32 ppk), indicating the potential application of $^{23}$Na-$^1$H-combined NMR for measuring brine salinity in porous materials in laboratory and in situ. Given the wide range of salinity in well production water from various formations, such fast and non-destructive salinity measurement can provide real-time salinity and also help determine the hydrogen index of the brine phase.

**Table 2. $^{23}$Na-$^{1}$H-combined quantification for Vycor glass and conventional rocks (mean ± uncertainty)[a]**

| Sample | porosity (p.u.)[b] | | | NMR salinity[c] (ppk) |
|---|---|---|---|---|
| | MD | $^{23}$Na | $^{1}$H | |
| Vycor #1 | 31.3 ± 0.3 | 32 ± 3 | 31.5 ± 0.1 | 32 ± 3 |
| Vycor #2 | 32.1 ± 0.3 | 36 ± 3 | 33.2 ± 0.1 | 35 ± 3 |
| Austin Chalk | 16.1 ± 0.2 | 17 ± 2 | 15.9 ± 0.3 | 34 ± 4 |
| Edward Limestone | 22.3 ± 0.2 | 22 ± 2 | 18.7 ± 0.3 | 37 ± 3 |

a) The uncertainty for MD results was calculated based on the systematic uncertainty with laboratory balance; the uncertainty for NMR results was derived from data fitting.

b) Column MD shows mass-difference porosity; columns $^{23}$Na and $^{1}$H show porosity derived from $^{23}$Na and $^{1}$H CPMG acquisitions, respectively.

c) NMR salinity was calculated using the mass of NaCl and H$_2$O, which was derived from $^{23}$Na and $^{1}$H CPMG acquisitions, respectively, following the definition:

$$\text{salinity} = \text{mass (NaCl)}/\text{mass (H}_2\text{O)}.$$

The $^1$H and $^{23}$Na NMR $T_2$ distributions for Austin Chalk, Berea 500, and Vycor glass are shown in Fig. 3. In the low-field $^1$H $T_2$ distributions (Fig. 3A), the difference in pore size between these samples is clear; Vycor glass has the shortest $T_2$ ($T_2$ peak at 20 ms; blue curve), Austin Chalk has longer $T_2$ (major $T_2$ peak at 30 ms; red curve), and Berea 500 has the longest $T_2$ (major $T_2$ peak at 500 ms; green curve). Vycor glass is synthesized with uniform 4-nm pores, resulting in a clear, single $^1$H $T_2$ peak. On the other hand, the conventional rocks have an inhomogeneous pore size distribution, reflected by more than one $T_2$ peak. The relaxation times can be converted to pore size ($S/V$) using Eq. 4 with knowledge of surface relaxivity (in this case, $\rho_2$). High-field $^1$H $T_2$ distributions (Fig. 3B) qualitatively match well with those from low-field experiments (Fig. 3A), with an obvious shift towards the short-$T_2$ direction. It can be noted that the high-field $T_2$ peaks uniformly decrease by four- to fivefold in comparison with the low-field peaks (high-field major $T_2$ peaks at 5 ms for Vycor glass, 6 ms for Austin Chalk, and 100 ms for Berea 500). The $T_2$ decrease is mainly due to diffusion in the presence of magnetic susceptibility induced internal magnetic field gradients.[34] This effect is inversely proportional to the square of both external magnetic-field strength and the internal magnetic-field gradients.[19]

For the conventional rocks, high-field $^{23}$Na $T_2$ distributions (Fig. 3C) are compared with the $^1$H $T_2$ distributions (Fig. 3B). In bulk NaCl solution, the $^1$H $T_2$ is on the order of seconds[26] whereas the $^{23}$Na $T_2$ is ~70 ms, under our experimental conditions (25° C and 9 T). The ratio $T_2(^1H):T_2(^{23}Na)$ in bulk solution is larger than that in these conventional rocks, due to different surface relaxivities ($\rho_2$'s) for $^1$H and $^{23}$Na (Eq. 4). The $T_2$ ratio of Austin Chalk versus Berea 500, calculated using the major $T_2$ peaks, is 1:20 in the $^{23}$Na domain and 1:17 in the $^1$H domain. Interestingly, bi-exponential

$^{23}$Na transverse relaxation is identified in Vycor glass (blue curve in Fig. 3C), which results from electric quadrupole coupling and will be explored in detail elsewhere.

The $^1$H $T_1$-$T_2$ maps for Austin Chalk, Berea 500, and Vycor glass are shown in Fig. 4. For the rocks, $T_1$-$T_2$ maps obtained at low (Fig. 4A) and high (Fig. 4B) fields demonstrate similar peak contour shapes but have different $T_1/T_2$ ratios. In the case of Austin Chalk and Berea 500, this is mainly due to the decrease in $T_2$ at high fields, which arises from diffusion in the susceptibility induced internal gradients, as discussed above. In the case of Vycor glass, the polar water molecules strongly interact with and are therefore slowed down by the hydrophilic pore surface, resulting in an increase in $T_1$, which causes higher $T_1/T_2$ ratio. The $T_1$-$T_2$ map for Vycor glass presents a single peak at both low (Fig. 4A) and high (Fig. 4B) fields, in contrast to the conventional rocks, reflecting its homogeneous pore size distribution.

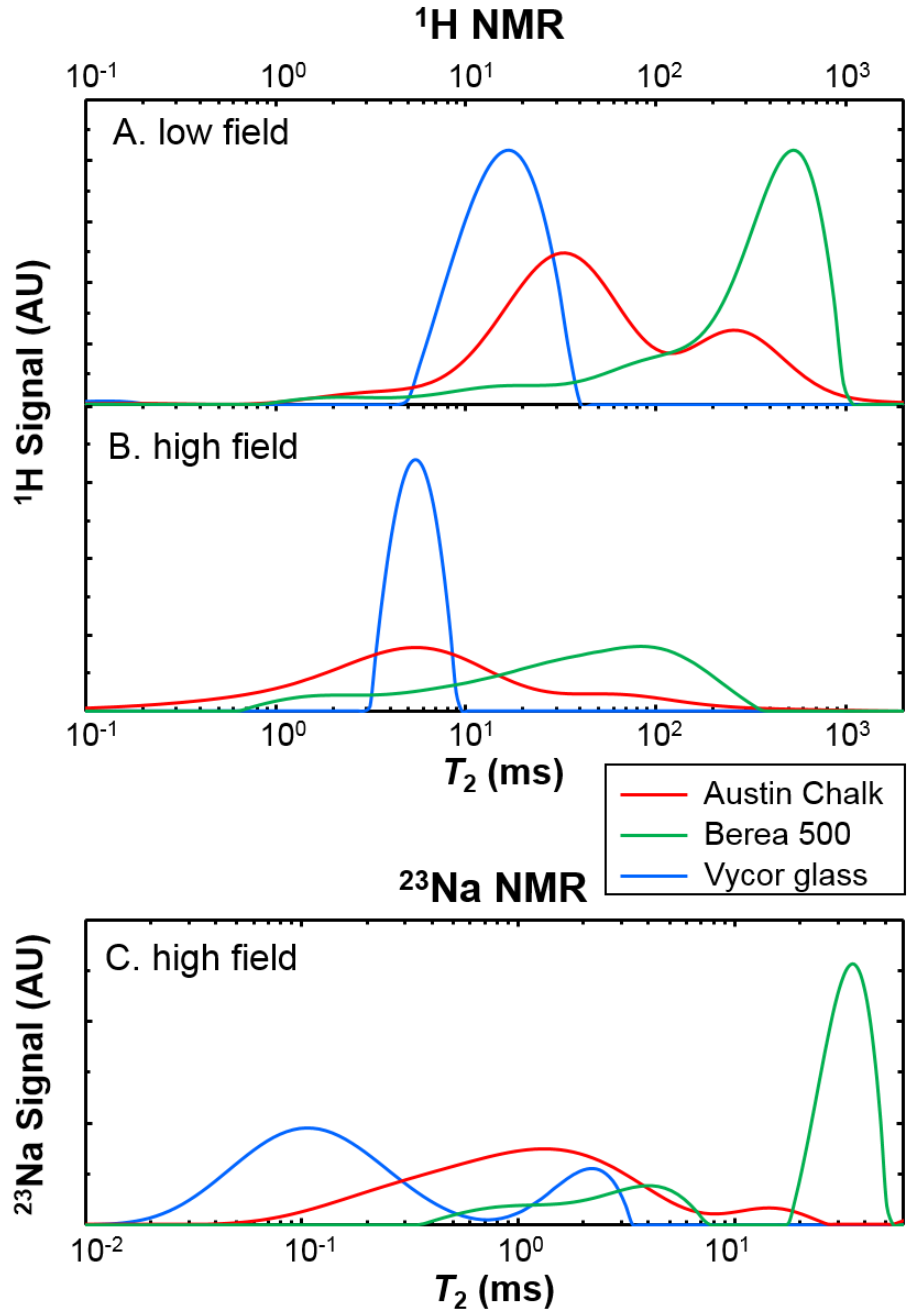

**Figure 3.** $T_2$ distributions in Vycor glass and conventional rocks with (A) $^1$H $T_2$ at low field (0.05 T), (B) $^1$H $T_2$ at high field (9 T), and (C) $^{23}$Na $T_2$ at high field (9 T). In each subplot, the $T_2$ distribution curves are colored as follows: Austin Chalk in red, Berea 500 in green, and Vycor glass in blue.

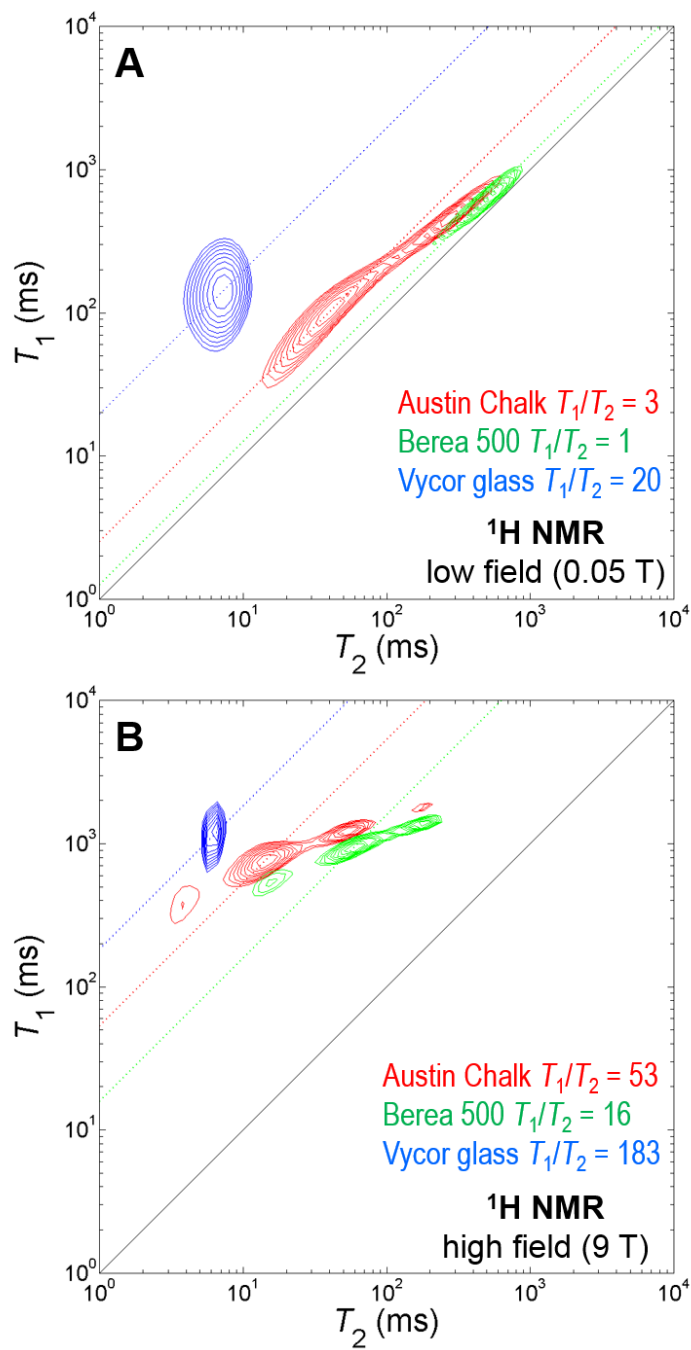

**Figure 4.** $T_1$-$T_2$ maps for brine-saturated Austin Chalk, Berea 500, and Vycor glass at (A) low and (B) high fields. Distribution contour lines are colored as follows: Austin Chalk in red, Berea 500 in green, and Vycor glass in blue. Dotted lines with corresponding color mark the $T_1/T_2$ ratio at the major peak in each sample.

## 4.2. Clays and Quartz

Samples of clay-brine mixtures were employed as models for studying the NMR response of clay-associated water. Comparisons were made with quartz-brine mixture, which is a proxy for the inorganic, non-clay porous networks found in rocks. NMR-based brine mass and salinity are derived from FID experiments. FID acquisition was used instead of CPMG to get around the fast transverse relaxation (short $T_2$) in both $^1$H and $^{23}$Na in clay-brine mixtures. The NMR-based quantification results are compared to the mass-difference results in Table 3, and they show good agreement. The quantification of $H_2O$ in the brine using $^1$H NMR matches much better with the mass-difference results for illite and bentonite in comparison with smectite.

High-field $T_2$ distributions for clay-brine and quartz-brine mixtures are shown in Fig. 5. The $T_2$ distribution in quartz is comparable to that observed in rocks such as Austin Chalk, whereas those in clays are about 100 us and thus shorter by one to two orders of magnitude. The clay $T_1$-$T_2$ distributions are single-valued for both $^1$H and $^{23}$Na (Figs. 6A and 8A). For the quartz-brine mixture, $^{23}$Na $T_1$-$T_2$ map (Fig. 8A) reveals two peaks, unlike that of $^1$H, reflecting the possible effect of quadrupole coupling. The $T_1$-$T_2$ peaks determined for isolated clays and quartz can be used for identifying clay-associated and free water components, as discussed in the next section.

**Table 3. $^{23}$Na-$^{1}$H-combined quantification for clays (mean ± uncertainty)[a]**

| Type of clay | | Illite | Bentonite | Smectite |
|---|---|---|---|---|
| m(NaCl) (mg) | $^{23}$Na | 2.4 ± 0.3 | 3.8 ± 0.1 | 2.6 ± 0.1 |
| | MD[b] | 2.51 ± 0.03 | 4.03 ± 0.04 | 2.49 ± 0.02 |
| m(H$_2$O) (mg) | $^{1}$H | 27.0 ± 2.0 | 47.0 ± 1.0 | 32.3 ± 0.4 |
| | MD[b] | 27.9 ± 0.3 | 44.7 ± 0.4 | 27.6 ± 0.3 |
| salinity[c] (ppk) | $^{23}$Na-$^{1}$H | 89 ± 18 | 81 ± 4 | 80 ± 4 |
| | MD[b] | 90.2 ± 0.2 | 90.2 ± 0.2 | 90.2 ± 0.2 |

a) The uncertainty for MD results was calculated based on the systematic uncertainty with laboratory balance; the uncertainty for NMR results was derived from data fitting.

b) Rows with MD show the quantities calculated based on mass-difference records measured using a laboratory balance when preparing the samples.

c) $^{23}$Na-$^{1}$H salinity was calculated using the mass of NaCl and H$_2$O, which was derived from $^{23}$Na and $^{1}$H FID acquisitions, respectively, following the definition:

$$\text{salinity} = \text{mass (NaCl)}/\text{mass (H}_2\text{O)}.$$

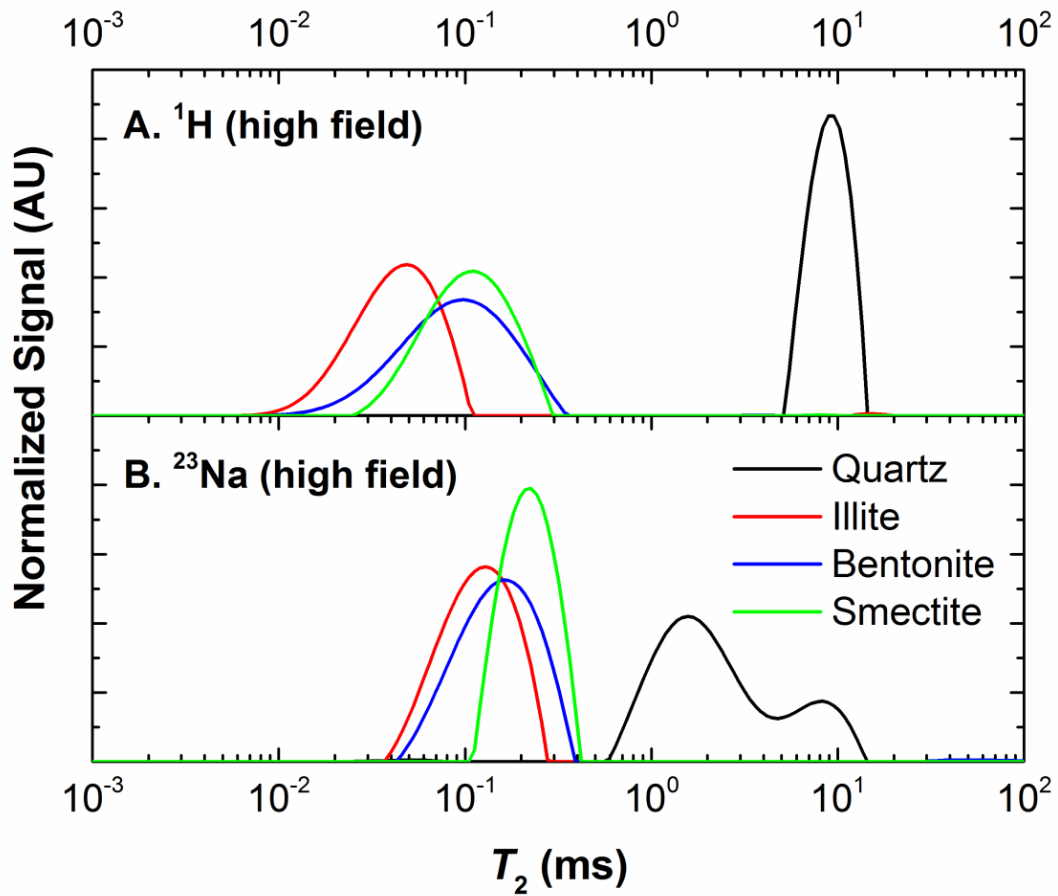

**Figure 5.** (A) $^1$H and (B) $^{23}$Na $T_2$ distributions in clay-brine and quartz-brine mixtures at high field (9 T). In each subplot, the $T_2$ distribution curves are colored as follows: quartz in black, illite in red, bentonite in blue, and smectite in green.

## 4.3. Shale

The $^1$H and $^{23}$Na $T_1$-$T_2$ maps for shale samples from three different depths are shown in Figs. 6 and 8, respectively, and the 1D $T_1$ and $T_2$ projections are shown separately in Fig. 9. The $^1$H $T_1$-$T_2$ maps for native-state samples are in subplots B1, C1, and D1 in Fig 6, and those for the corresponding brine-saturated samples are in subplots B2, C2, and D2. By comparing the native-state and brine-saturated maps, the locations of signal increase in brine-saturated samples can be identified; these are marked by two sets of boxes. The $^1$H subpopulation in the solid-line box is associated with $T_2$ ranging from 0.1–0.5 ms and $T_1$ of ~10 ms, which corresponds to water interacting with clays, as suggested by the $T_1$-$T_2$ map for clay-brine mixtures in Fig. 6A. The second $^1$H subpopulation, marked by the dashed-line box, spreads across the $T_2$ range of 1–10 ms with $T_1$ at ~100–200 ms, corresponding to free fluids in the inorganic porosity, as shown by the $T_1$-$T_2$ peak of quartz-brine mixture (Fig. 6A). According to these results, a cutoff in the $^1$H-$T_1$ domain ($T_1$ = 50 ms; dashed lines) can be established for separating the clay-associated water from the other components such as bitumen and kerogen. The $^1$H $T_1$-$T_2$ peak distributions for all fluid components are illustrated in Fig. 7.

No $^{23}$Na signal was observed in the native-state shale, so only the $^{23}$Na NMR results from the saturated samples are shown (1D $T_1$ in Fig. 9C, 1D $T_2$ in Fig. 9D, and 2D $T_1$-$T_2$ in Fig. 8B–8D). Three $T_2$ peaks (Fig. 9D) were observed in all shale samples. The peaks with medium and long $T_2$ (on the order of 1 and 10 ms, respectively) are similar to those observed in brine-saturated quartz (Fig. 5B), indicating free water components. The peak with short $T_2$ (~0.1 ms) is comparable to that found in clay-brine mixtures (Fig. 5B), arising from clay-associated water. This peak assignment is also supported by comparing $T_1$-$T_2$ maps of brine-saturated shale (Fig. 8B–8D) with that of clay and quartz (Fig. 8A). Using $^{23}$Na $T_1$-$T_2$ maps, a cutoff in the $T_2$ domain ($T_2 = 0.4$ ms; dashed lines) can be identified between bound and free brine. Importantly, these $^{23}$Na peaks are well separated in both 1D and 2D plots, making the fluid differentiation feasible.

Based on 1D relaxation distributions, a $^1$H signal increase in the brine-saturated shale can be observed along both $T_1$ (Fig. 9A) and $T_2$ (Fig. 9B) dimensions. Particularly in Fig. 9B, the signal increase appears in the region with $T_2 > 0.1$ ms. This range of $T_2$ corresponds to the location of multiple components like bound oil, movable oil, bound water, and free water, making fluid typing via 1D relaxometry alone challenging. The major $T_2$ peak in the native-state shale sample (Fig. 9B) is at ~50 μs, which corresponds to immobile components (e.g., kerogen and bitumen) and the clay-associated water. NMR signals from the solid and viscous hydrocarbon fractions (kerogen and bitumen) are largely overlapped with signals from bound water, which causes a lack of contrast for separating these hydrocarbon and water components solely based on the 1D $T_2$ method.

Brine porosity results are presented in Table 4. According to mass difference calculations, only ~0.1–0.5 mg NaCl was introduced into each shale sample following the brine saturation procedure. The mass-difference porosity is 1–3 p.u. (MD row in Table 4), implying small $^{23}$Na NMR signals to be observed in these samples. The MD results in Table 4 reflect the mass readings recorded before NMR measurements. The $^{23}$Na CPMG acquisitions were used and $^{23}$Na-NMR-based porosities agree well with the MD results in samples 1 and 2. The porosity estimate for shale sample 3 is not listed. After the pressurized saturation procedure, some pieces from sample 3 were too small and soft to be transferred into the laboratory balance, so the MD porosity was not calculated. The $^{23}$Na porosity for sample 3 is $1.8 \pm 0.2$ p.u., which aligns well with the other two samples. Determination of $^1$H-based porosity and salinity is challenged by the presence of strong residual $^1$H signal in the native-state sample as well as the observable water loss through evaporation during the early stage of NMR experiments. The $^{23}$Na NMR experiments were less affected by these factors, especially because there was no background $^{23}$Na signal in the native-state shale. $^{23}$Na NMR is herein validated as a promising method for quantitatively analyzing the brine content in shale. Further optimization of the current sample preparation protocol, which should aim at improving water evaporation control during NMR experiments, may facilitate the incorporation of quantitative $^1$H measurements and thus the determination of component-specific salinity in the brine-saturated shale. The applications and advantages of high-field $^1$H and $^{23}$Na NMR methods are summarized in Table 5 and compared to low-field measurements.

**Table 4. Brine porosity estimation in shale using $^{23}$Na NMR (mean ± uncertainty)[a]**

| Sample | | 1 | 2 |
|---|---|---|---|
| depth (ft) | | xx87 | xx98 |
| porosity (p.u.) | $^{23}$Na | 2.6 ± 0.2 | 1.5 ± 0.1 |
| | MD[b] | 2.87 ± 0.04 | 1.41 ± 0.04 |

a) The uncertainty for MD results was calculated based on the systematic uncertainty with laboratory balance; the uncertainty for NMR results was derived from data fitting.

b) Rows with MD show the porosity calculated based on mass-difference records measured using laboratory balance when preparing the samples.

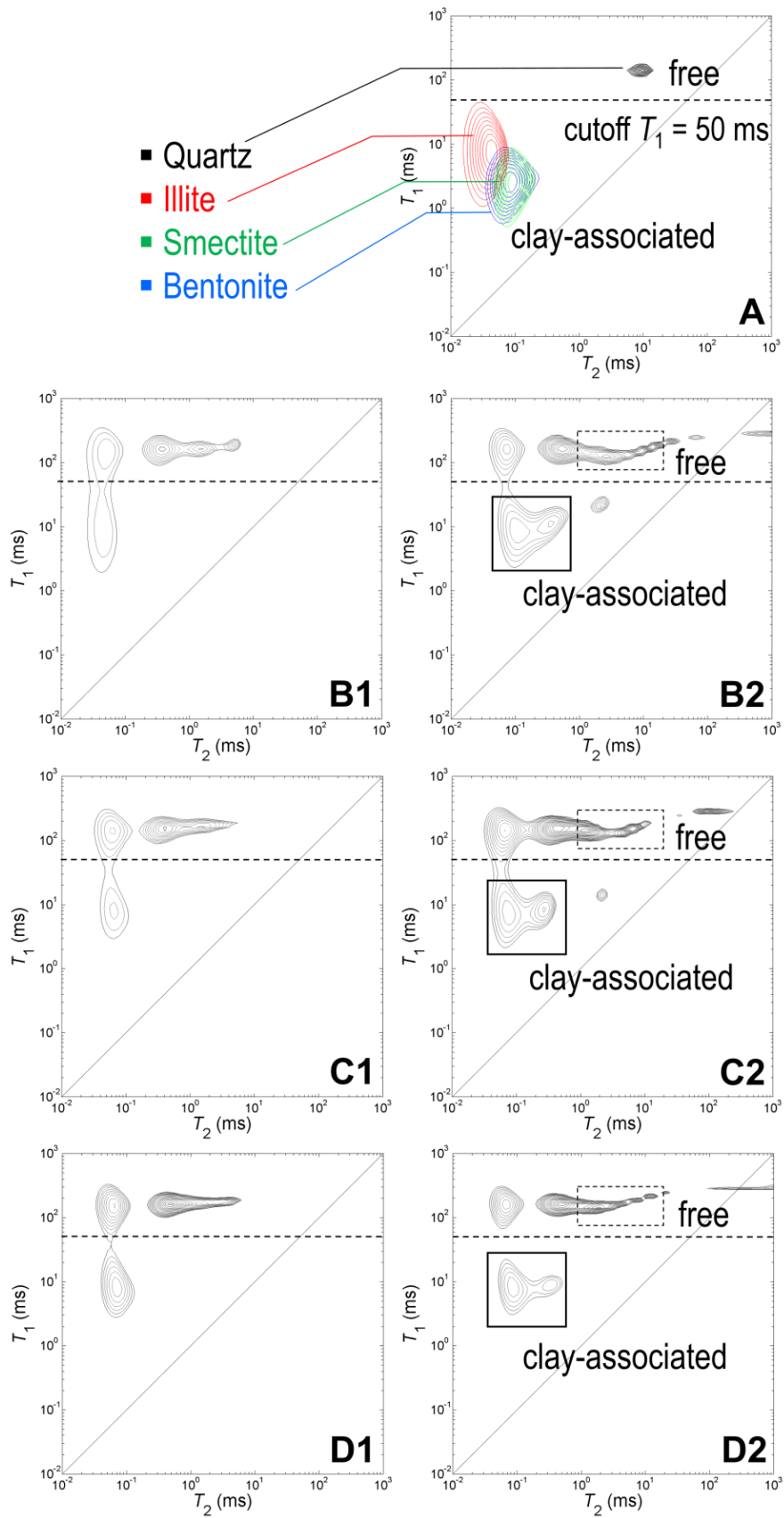

**Figure 6.** $^1$H $T_1$-$T_2$ maps for clays, quartz, and shale. (A) is the $^1$H $T_1$-$T_2$ map for illite (red contour lines), smectite (green contour lines), bentonite (blue contour lines), and quartz (black contour lines), which were saturated with brine. (B) through (D) are the $^1$H $T_1$-$T_2$ maps for Eagle Ford shale samples 1 through 3, respectively, where (B1), (C1), and (D1) are from native-state samples, and (B2), (C2), and (D2) are from brine-saturated samples. A cutoff $T_1$ of 50 ms (dashed line in all subplots) is established to differentiate clay-associated and free water components. For brine-saturated shale samples, (B2), (C2), and (D2), laboratory-introduced brine components are marked by two sets of boxes; solid-line boxes show the location of clay-associated water, and dashed-line boxes show that of free water. These boxes are drawn according to a qualitative comparison between the $T_1$-$T_2$ maps of native samples and those of saturated ones.

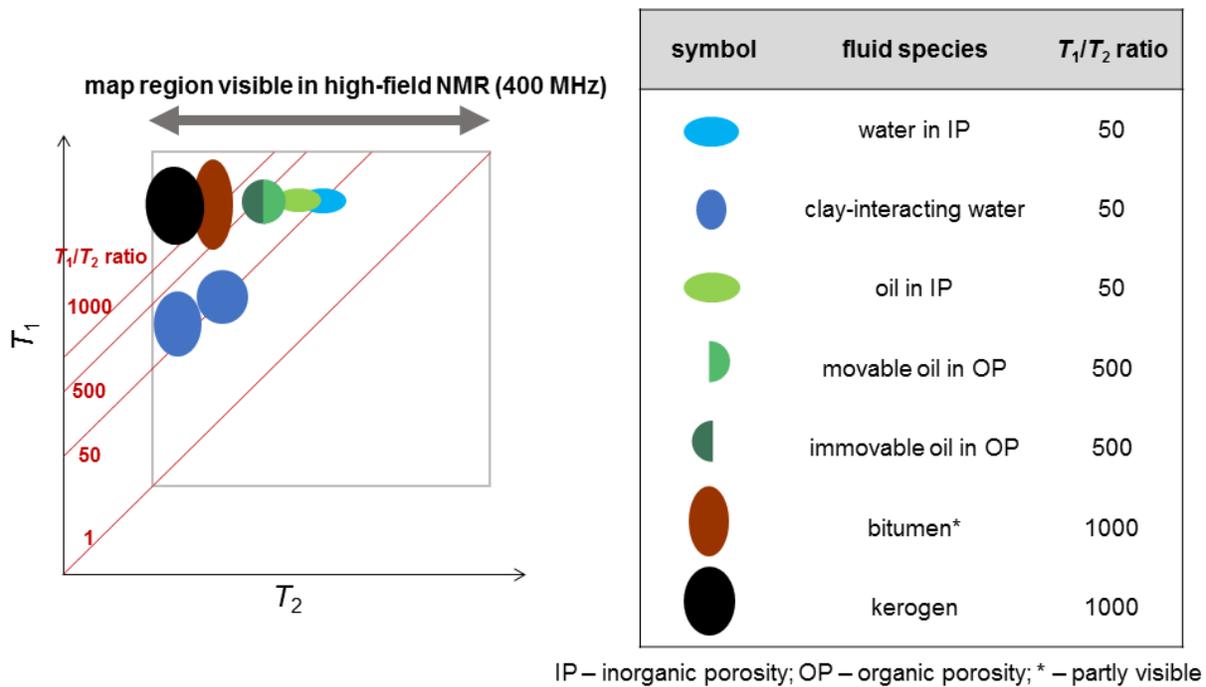

**Figure 7.** Illustrative high-field (400 MHz) $^1$H $T_1$-$T_2$ map for fluid components in Eagle Ford shale.

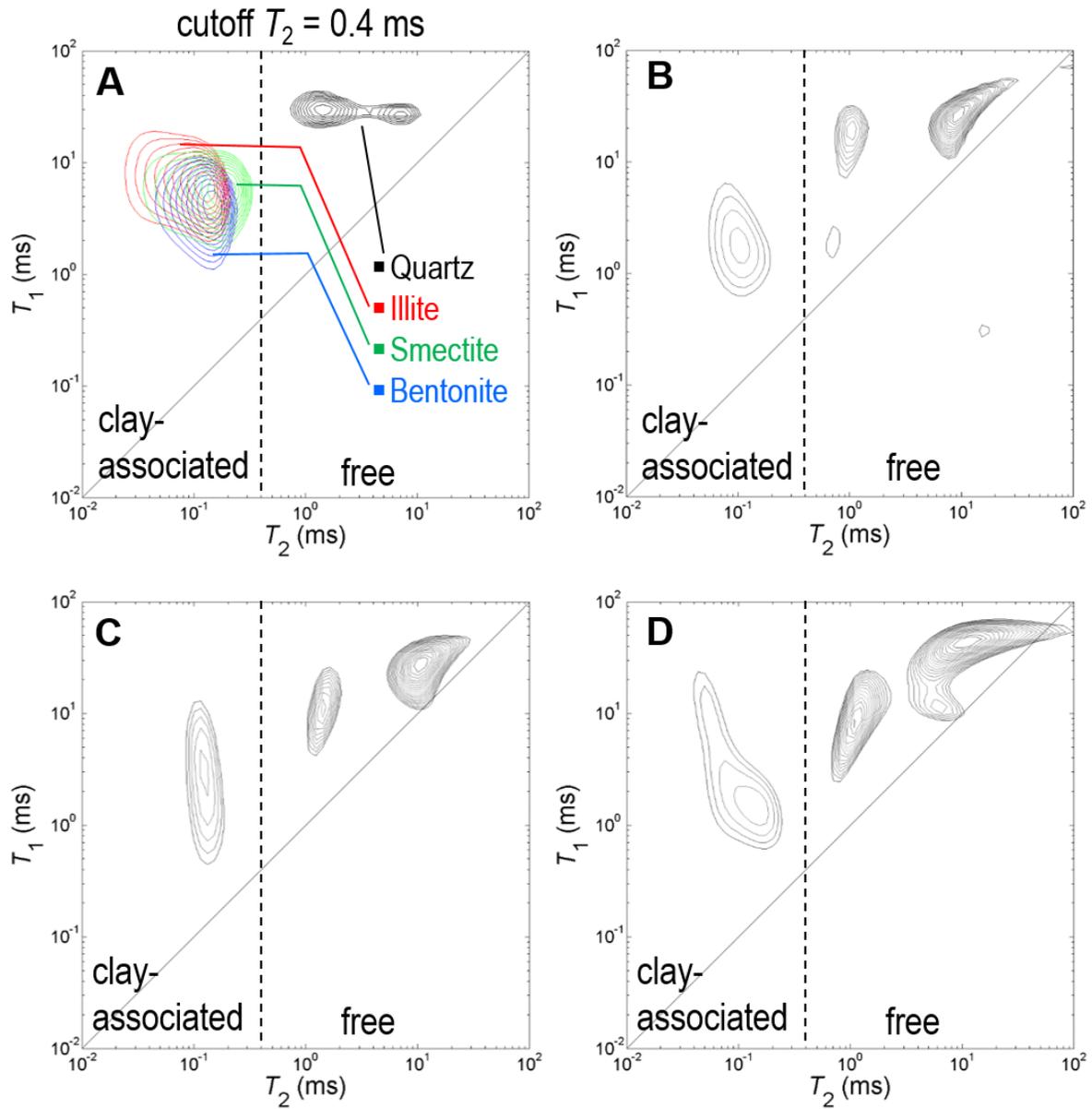

**Figure 8.** $^{23}$Na $T_1$-$T_2$ maps for clays, quartz, and shale. (A) is the $^{23}$Na $T_1$-$T_2$ map for illite (red contour lines), smectite (green contour lines), bentonite (blue contour lines), and quartz (black contour lines), which were saturated with brine. (B) through (D) are the $^{23}$Na $T_1$-$T_2$ maps for brine-saturated Eagle Ford shale samples 1 through 3, respectively. A cutoff $T_2$ of 0.4 ms (dashed line in all subplots) is established to differentiate clay-associated and free water components.

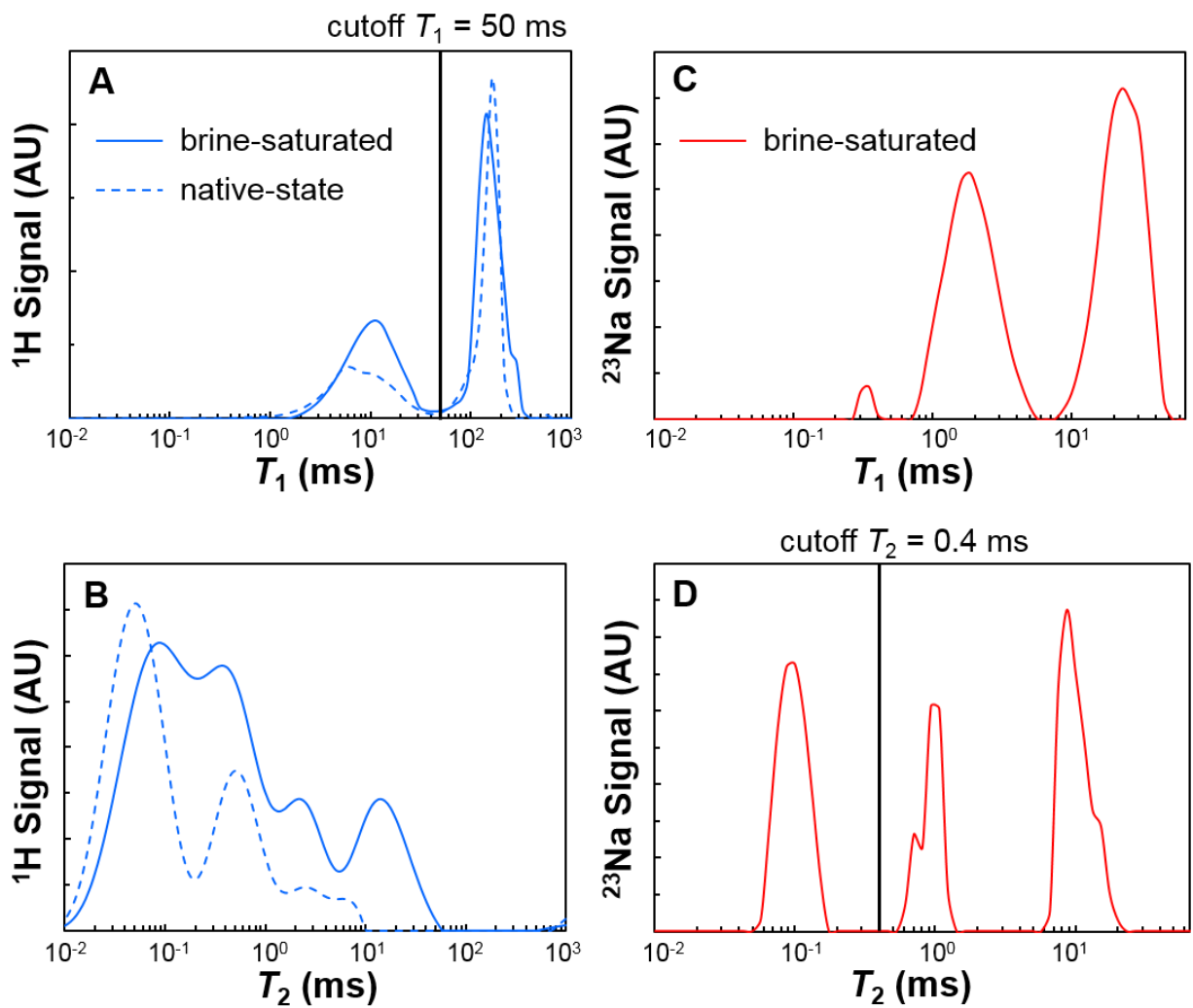

**Figure 9.** Example 1D $T_1$ and $T_2$ distribution for shale (Eagle Ford shale sample 1). (A) and (B) are $^1$H $T_1$ and $T_2$ distribution, respectively, in shale. Distribution curves for the brine-saturated sample are shown in solid blue lines, and those for the native-state sample are shown in dashed blue lines. (C) and (D) are $^{23}$Na $T_1$ and $T_2$ distribution, respectively, in the brine-saturated sample. The cutoff $T_1$ for $^1$H NMR and cutoff $T_2$ for $^{23}$Na NMR are derived from Figs. 6 and 8, respectively.

**Table 5. Applications of low and high field NMR methods in various types of samples**

| | | Low field (≤ 0.05 T)[20, 22] | High field (> 0.25 T) |
|---|---|---|---|
| Conventional rocks | 1D relaxometry | *$^1$H NMR:*<br>• Pore size distribution<br>• Total hydrocarbon and water filled porosity | *$^1$H-$^{23}$Na NMR:*<br>• Separates hydrocarbon and brine porosities; when sodium salinity (concentration) is known |
| Unconventional rocks | 1D relaxometry | *$^1$H NMR:*<br>• Sensitive to part of bitumen<br>• Insensitive to kerogen<br>• Sensitive to bound water, light hydrocarbon and the free fluids | *$^1$H NMR:*<br>• Sensitive to bitumen<br>• Sensitive to kerogen<br>• Sensitive to bound water, light hydrocarbon and the free fluids<br><br>*$^{23}$Na NMR:*<br>• Separates clay-associated and free brine |
| | 2D $T_1$-$T_2$ | *$^1$H NMR:*<br>• Identifies bitumen<br>• Separates oils in organic and inorganic porosities<br>• No contrast for free fluids | *$^1$H NMR:*<br>• Separates clay-associated water from hydrocarbons<br>• Low contrast for free fluids<br><br>*$^{23}$Na NMR:*<br>• Separates clay-associated and free brine<br><br>*$^1$H-$^{23}$Na NMR:*<br>• Quantifies clay-associated brine, free brine, solid and viscous hydrocarbons (kerogen and bitumen), and free hydrocarbons; when sodium salinity (concentration) is known |

## 5. CONCLUSIONS

In this study, $^{23}$Na-$^{1}$H-combined, high-field 2D NMR relaxometry is demonstrated to be a quantitative method for separately characterizing the brine and hydrocarbon contents in various porous materials, from porous glass, to conventional rocks, clays, and shale. We show that NMR $T_1$-$T_2$ relaxometry at high Larmor frequencies can separate the immobile hydrocarbons such as kerogen and bitumen from the bound water, unlike in the case of low Larmor frequency measurements. We also show that $^{23}$Na NMR is capable of quantifying small amount (a few mg) of NaCl in brine and therefore can measure the brine-filled porosity when the NaCl salinity is known. The application of $^{23}$Na-$^{1}$H-combined NMR can therefore provide superior fluid typing and salinity information in unconventional shale, which is presently a great challenge in formation evaluation. The laboratory characterizations provide the basis for in situ $^{23}$Na and $^{1}$H NMR downhole logging measurements, wherein brine salinity, hydrocarbon content and total water-filled porosity can be determined, potentially playing a critical role for NMR logging in shale.

AUTHOR INFORMATION

**Corresponding Author**

*R.K.: E-mail: RViswanathan@slb.com

**ACKNOWLEDGMENT**

Donghan Yang would like to acknowledge the support of Schlumberger-Doll Research Center through the internship program during which this work was done. The authors would like to thank Martin D. Hürlimann and Yi-Qiao Song for fruitful discussions, Lalitha Venkataramanan for helpful feedback on the manuscript, Kamilla Fellah for help with sample preparation and Schlumberger for permitting us to publish the results.

REFERENCES

1.     Kausik, R.; Craddock, P. R.; Reeder, S. L.; Kleinberg, R. L.; Pomerantz, A. E.; Shray, F.; Lewis, R. E.; Rylander, E., Novel Reservoir Quality Indices for Tight Oil. In *Unconventional Resources Technology Conference*, Society of Petroleum Engineers: San Antonio, Texas, USA, 2015.

2.     Watson, A. T.; Chang, C. T. P., Characterizing porous media with NMR methods. *Prog Nucl Mag Res Sp* **1997,** 31, 343-386.

3.     Barrie, P. J., Characterization of porous media using NMR methods. In *Annual Reports on NMR Spectroscopy*, Academic Press: 2000; Vol. Volume 41, pp 265-316.

4.     Song, Y. Q.; Cho, H.; Hopper, T.; Pomerantz, A. E.; Sun, P. Z., Magnetic resonance in porous media: Recent progress. *J Chem Phys* **2008,** 128, (5).

5.     Cho, H. J.; Sigmund, E. E.; Song, Y. Q., Magnetic Resonance Characterization of Porous Media Using Diffusion through Internal Magnetic Fields. *Materials* **2012,** 5, (4), 590-616.

6.     Song, Y.-Q.; Ryu, S.; Sen, P. N., Determining multiple length scales in rocks. *Nature* **2000,** 406, (6792), 178-181.

7.     Kleinberg, R. L.; Jackson, J. A., An introduction to the history of NMR well logging. *Concept Magnetic Res* **2001,** 13, (6), 340-342.

8.     Newling, B., "Gas flow measurements by NMR". *Prog Nucl Mag Res Sp* **2008,** 52, (1), 31-48.

9.     Topgaard, D., Probing biological tissue microstructure with magnetic resonance diffusion techniques. *Curr Opin Colloid In* **2006,** 11, (1), 7-12.

10. Kausik, R.; Cao Minh, C.; Zielinski, L.; Vissapragada, B.; Akkurt, R.; Song, Y.-Q.; Liu, C.; Jones, S.; Blair, E., Characterization of Gas Dynamics in Kerogen Nanopores by NMR. In *SPE Annual Technical Conference and Exhibition*, Society of Petroleum Engineers: Denver, Colorado, USA, 2011.

11. Wang, H. J.; Mutina, A.; Kausik, R., High-Field Nuclear Magnetic Resonance Observation of Gas Shale Fracturing by Methane Gas. *Energ Fuel* **2014,** 28, (6), 3638-3644.

12. Papaioannou, A.; Kausik, R., Methane Storage in Nanoporous Media as Observed via High-Field NMR Relaxometry. *Phys Rev Appl* **2015,** 4, (2).

13. Brown, R. J. S.; Gamson, B. W., Nuclear magnetism logging. *Transactions of the American Institute of Mining and Metallurgical Engineers* **1960,** 219, (8), 199-207.

14. Timur, A., Pulsed Nuclear Magnetic Resonance Studies of Porosity Movable Fluid and Permeability of Sandstones. *SPE-0412-0034-JPT* **1969,** 21, (Jun), 775-&.

15. Loren, J. D.; Robinson, J. D., Relations between Pore Size Fluid and Matrix Properties, and NML Measurements. *Soc Petrol Eng J* **1970,** 10, (3), 268-278.

16. Robinson, J. D.; Loren, J. D.; Vajnar, E. A.; Hartman, D. E., Determining Residual Oil with Nuclear Magnetism Log. *SPE-0412-0034-JPT* **1974,** 26, (Feb), 226-236.

17. Neuman, C. H.; Brown, R. J. S., Applications of Nuclear Magnetism Logging to Formation Evaluation. *SPE-0412-0034-JPT* **1982,** 34, (12), 2853-2862.

18. Kenyon, W. E.; Day, P. I.; Straley, C.; Willemsen, J. F., A Three-Part Study of NMR Longitudinal Relaxation Properties of Water-Saturated Sandstones. **1988**.

19. Vinegar, H.; Kleinberg, R., NMR properties of reservoir fluids. *The Log Analyst* **1996**, (6), 20.
38

20. Kausik, R.; Fellah, K.; Rylander, E.; Singer, P. M.; Lewis, R. E.; Sinclair, S., NMR Petrophysics for Tight Oil Shale Enabled by Core Resaturation. In *International Symposium of the Society of Core Analysts*, Avignon, France, 2014; Vol. 73.

21. Hu, H. T.; Xiao, L. Z.; Wu, X. L., Corrections for downhole NMR logging. *Petrol Sci* **2012,** 9, (1), 46-52.

22. Kausik, R.; Fellah, K.; Rylander, E.; Singer, P. M.; Lewis, R. E.; Sinclair, S. M., NMR Relaxometry in Shale and Implications for Logging. In *SPWLA 56th Annual Logging Symposium*, Society of Petrophysicists and Well-Log Analysts: Long Beach, California, USA, 2015.

23. Godefroy, S.; Korb, J.-P.; Fleury, M.; Bryant, R., Surface nuclear magnetic relaxation and dynamics of water and oil in macroporous media. *Physical Review E* **2001,** 64, (2), 021605.

24. Ferris, J.; Tutunjian, P.; Vinegar, H., Nuclear magnetic resonance imaging of sodium-23 in cores. *The Log Analyst* **1993,** 34, (03).

25. Eisensta, M.; Friedman, H. L., Nuclear Magnetic Relaxation in Ionic Solution .I. Relaxation of $^{23}$Na in Aqueous Solutions Of NaCl and NaClO$_4$. *J Chem Phys* **1966,** 44, (4), 1407.

26. Chary, K. V. R.; Govil, G., *NMR in Biological Systems: From Molecules to Human*. Springer: 2008.

27. Levitt, M. H., *Spin Dynamics: Basics of Nuclear Magnetic Resonance*. Wiley: 2001.

28. Woessner, D. E., NMR relaxation of spin-(3)/(2) nuclei: Effects of structure, order, and dynamics in aqueous heterogeneous systems. *Concept Magnetic Res* **2001,** 13, (5), 294-325.

29. Rijniers, L. A.; Magusin, P. C. M. M.; Huinink, H. P.; Pel, L.; Kopinga, K., Sodium NMR relaxation in porous materials. *J Magn Reson* **2004,** 167, (1), 25-30.

30. Porion, P.; Faugere, M. P.; Lecolier, E.; Gherardi, B.; Delville, A., Na-23 nuclear quadrupolar relaxation as a probe of the microstructure and dynamics of aqueous clay dispersions: An application to Laponite gels. *J Phys Chem B* **1998,** 102, (18), 3477-3485.

31. Porion, P.; Al Mukhtar, M.; Meyer, S.; Faugere, A. M.; van der Maarel, J. R. C.; Delville, A., Nematic ordering of suspensions of charged anisotropic colloids detected by Na-23 nuclear magnetic resonance. *J Phys Chem B* **2001,** 105, (43), 10505-10514.

32. Headley, L. C., Nuclear Magnetic-Resonance Relaxation of Na-23 in Porous Media Containing NaCl Solution. *J Appl Phys* **1973,** 44, (7), 3118-3121.

33. Song, Y. Q.; Venkataramanan, L.; Hurlimann, M. D.; Flaum, M.; Frulla, P.; Straley, C., T-1-T-2 correlation spectra obtained using a fast two-dimensional Laplace inversion. *J Magn Reson* **2002,** 154, (2), 261-268.

34. Stejskal, E. O.; Tanner, J. E., Spin Diffusion Measurements: Spin Echoes in the Presence of a Time-Dependent Field Gradient. *J Chem Phys* **1965,** 42, (1), 288-292.